# Distinguishing Bulk-Diffusion from Surface-Desorption Limited Gas Release Processes

Ricardo E. Avila

Departamento de Materiales Nucleares, Comisión Chilena de Energía Nuclear,
Cas. 188-D, Santiago, Chile (ravila@cchen.cl)

ABSTRACT

The release of a gas limited by surface desorption, or by diffusion from the bulk of spherical pebbles is revisited. A method is proposed to identify the release limiting process, by comparing a partial temperature ramp, up to slightly beyond the release peak, followed by a rapid temperature drop, to a second, full release ramp. Comparing the release curve from the second ramp to that of the first one: i) the peak is unmoved in first order desorption kinetics, and moves to higher temperature in the other cases, ii) as compared to the Arrhenius analysis of the first curve, that of the second is, again, identical in first order kinetics, in second order desorption it maintains the slope but lowers the intercept at the reciprocal temperature origin, and it is inapplicable in bulk diffusion kinetics.

The release of a gas species, which is initially distributed throughout the bulk of a solid, is controlled mostly by the slower of either bulk diffusion or surface desorption. Both processes have been studied extensively[1,2]. The present work builds on the well known analytic results for systems where only one process of either diffusion or desorption controls the gas release, and it is applied to homogeneous spherical pebbles. The results will shed additional light, also, towards identifying the order of surface-only desorption.

The surface vs. bulk control problem has been approached by Bertone[3] using isothermal release of tritium from $Li_2O$. Drawbacks of that approach are discussed in the present article, and a complementary experiment is proposed. The presentation will refer to diffusion and desorption of a gas to the surrounding atmosphere, however, the ambient needs not be a gas, an agitated liquid being just as appropriate.

The Redhead[4] model for surface desorption proposes the evolution of the coverage (fraction of surface desorption sites occupied) remaining by time $t$, as

$$\frac{df}{dt} = -k(t) f^n, \qquad (1)$$

where $k(t)$ is the temperature-activated release coefficient,

$$k(t) = k_{os} \exp(-E_{as}/k_B T), \qquad (2)$$

where, $k_{os}$ is the pre-exponential factor, $E_{as}$ is the activation energy, $k_B$ is the Boltzmann constant, and $T$ is the absolute temperature. In this model, the possible coverage, and temperature dependences of $k_{os}$ are disregarded.

Equation (1) has been solved for the modeling of desorption, as well as for thermoluminescence[5] yields, assuming some specific temperature schedules. A generalization is obtained for arbitrary time dependence of $k(t)$ by introducing the Kirchhoff transformation[1]

$$\Lambda(t) = \int_0^t k(t')\,dt'. \qquad (3)$$

Thus, the general solution of eq. (1), for order 1, is

$$f(\Lambda) = f_o \exp(-\Lambda) \qquad (4)$$

where $f_o$ is the initial value of $f$. For a general real value of $n > 1$,

$$f(\Lambda) = \left[ f_o^{1-n} - (1-n)\Lambda \right]^{1/(1-n)}. \qquad (5)$$

These solutions are applicable to the determination of the kinetic parameters by either nonlinear least squares adjustment, or by Arrhenius analysis, as done[6] in the mathematically equivalent study of thermoluminescence yields, or regarding the coverage dependence of the desorption constants[7, 8], or surface desorption kinetics[9]. In summary, the Arrhenius analysis is obtained by writing the surface coverage as $f = f_o F/F_o$, where $F$ is the number of molecules on the surface, which contains $N_o$ adsorption sites, $F_o = F(t=0)$, and $f_o = F_o/N_o$ is the initial surface coverage. An experiment with a calibrated sensor provides the $dF/dt$ release curve, from which $F_o$ and $F(t)$ follow by integration from 0, or $t$, respectively, to the high temperature end of the experiment. Then, from eqs. (1) and (2), or by solving eq. (5) for $\Lambda$, and in slightly greater detail than in the aforementioned references,

$$f_o^{n-1} k_{os} \exp(-E_{as}/k_B T) = -\frac{1}{F_o}\frac{dF}{dt} \bigg/ (F/F_o)^n, \qquad (6)$$

from which the Arrhenius plot provides $E_{as}/k_B$ as the slope, and $\ln(f_o^{n-1} k_{os})$ as the $1/T \to 0$ intercept. The latter value is just $\ln(k_{os})$, for $n = 1$, however, for $n > 1$, extracting the pre-exponential factor, $k_{os}$, of the desorption constant requires a determination of $f_o$. For simple, truly surface processes, the latter may be assured to be unit by a saturating exposure, but it may be quite uncertain where the gas species may migrate into and out of the bulk, since, in that case, the relevant number of ad- or ab-sorbate sites may include a large number of bulk sites. This uncertainty does not affect the slope of the graph, but it translates into a possibly large offset to $f_o^{n-1} k_{os}$.

This analysis applies to a strictly surface desorption process, or to release from the bulk, where diffusion proceeds much faster that surface desorption.

If the gas is initially distributed throughout the bulk of a solid, it must diffuse to the surface to desorb. Here, the slower of the two processes, diffusion or desorption, controls the overall process. In the case of surface desorption control, eq. (1) can be used, after a reinterpretation of the meaning of $f$, considering the partition of the gas into the bulk and surface distributed portions. However, if the SD process is fast, leaving an inefficient bulk diffusion process to control the overall process, the diffusion equation must be solved.

The gas release limited by bulk diffusion will be calculated by assuming the gas species to be initially homogeneously distributed throught the interior (bulk) of homogeneous spherical pebbles. After normalization to unit total gas charge, the gas charge released up to time $t$ is [Carslaw and Jaeger[1], eq. (9.3.8)]

$$q(\lambda) = 1 - \frac{6}{\pi^2} \sum_{n=1}^{\infty} \frac{1}{n^2} \exp\left[-n^2 \pi^2 \lambda\right]. \qquad (7)$$

where $\lambda(t) = (k_{ob}/a^2)\Lambda(t)$, $a$ is the pebble radius, and $\Lambda(t)$ is obtained from eq. (3) with $D(t)$ in place of $k(t)$, $D(t)$ being the diffusion coefficient, assumed to be concentration independent, and given as $k(t)$ [eq. (2)], with $k_{ob}$ and $E_{ab}$ in place of $k_{os}$ and $E_{as}$.

A method[10], based on eq. (7), and akin to King's method [eq. (6)], has been devised towards extracting the diffusion kinetics from single-peak gas release curves, and will be used, in the next section to test various assumed release kinetics.

Where bulk diffusion and surface desorption may proceed at similar rates, a method to distinguish one from the other has been proposed by Bertone[3] [beware of a few misprints at and following his eq. (10)]. The method is applicable to surface release of first order only, and it may fail if the detector is contaminated by the gas released (as in monitoring tritium release with ionization chambers), since the distinction relies on a rapidly diminishing release trace.

A more robust approach is proposed, now, consisting of i) an initial upward temperature ramp which goes slightly past the release peak, followed by ii) a rapid temperature drop to a negligible release rate, to finish with iii) a second upwards ramp which completes the gas release.

A calculation for an SD-2 process, following that ramp sequence, is shown in Fig. 1, with $f_o = 1$, $k_{os} = 10^{13}$ / s, and $E_{as} = 2.21$ eV (50.9 kcal/mol). The upwards ramps are applied at 5 °C/min. The exponential temperature drop starts where the release rate has dropped by 5% past the maximum, at $T_d = 458.9$ °C.

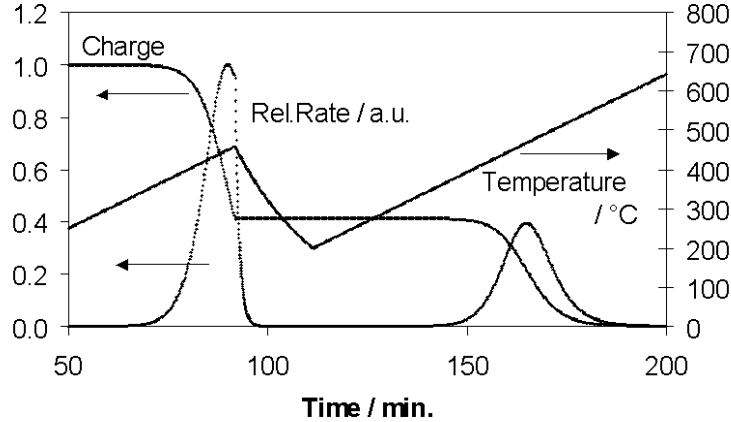

**Fig. 1.** Second order surface desorption from a surface initially at full coverage. The first linear ramp peaks at 450 °C, and the second one peaks at 470.3 °C.

Arrhenius analyses, assuming SD-1, SD-n (n>1), and BLR kinetics, are, then, applied to the two upwards ramps release data. Here, the charge normalization [$F_o$ in eq. (6)] for the first ramp is carried out over the whole process. Then, towards the analyses of the second ramp, the integrations for $\Lambda(t)$ and $F_o$ start at the onset of the second ramp, as if this were the only release ramp.

These analyses, as applied to synthetic data generated assuming an SD-2 model, are depicted in Fig. 2. The salient observations from that figure, and of the plots resulting from application of the three models to synthetic data generated with SD-1 and BLR models, are:

1. In SD-1 kinetics, since the only effect of the first, partial up-ramp, on the second one is a diminished total charge, each analysis (SD-1, SD-2, and BLR) of the second ramp is superimposed on that of the first ramp. SD-1 kinetics is the only process displaying this behavior.
2. In contrast, the SD-1 and SD-2 analyses of an SD-2 process (as in Fig. 2) yield curves which approach each other at the low temperature side, whereas this happens at the high temperature end when these analyses are applied to BLR data.
3. In all cases, SD-2 analysis is clearly distinct from SD-1 and BLR analyses, and the last two are closer (in the sense that they yield close to straight lines, if observed over narrow temperature ranges, when applied to the non-matching data), which may lead to misinterpretations.

These observations are summarized in Table I.

An additional distinction between SD-1, SD-2 and BLR processes follows from the upward temperature shift of the second or later ramp peaks, from that of the first ramp. Figure 3 shows the dependence of the peak temperature of SD-1, SD-2, and BLR processes, on the initial coverage (of SD kinetics) or on the fraction of the initial charge (for the BLR case) at the onset of the second or later ramps.

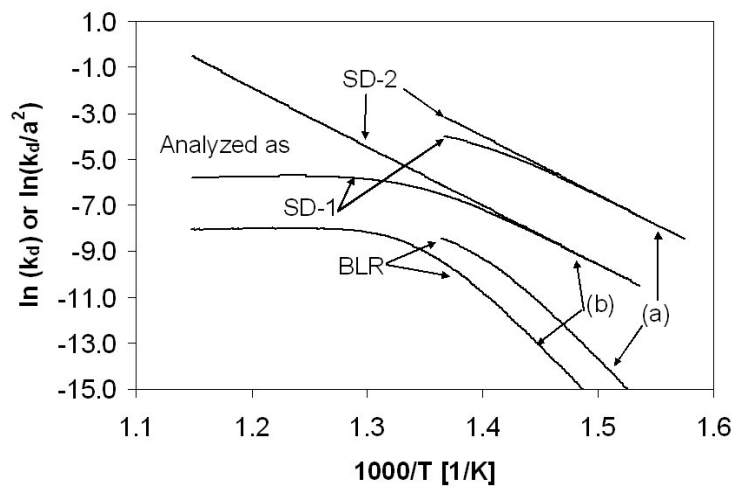

**Fig. 2**. Second order desorption data analyzed according to each of the models, as indicated, for the first (a), and second (b) ramps. The ordinate axis is $\ln(k_o)$ for SD analysis, and $\ln(k_o/a^2)$ for BLR analysis, with time in seconds and length in cm. The SD-1 and SD-2 analysis of the first ramp are displaced upwards by 2 for clarity.

As expected, the peak temperature of an SD-1 process is independent of the initial coverage. For a BLR process the peak temperature approaches the value corresponding to an initial bulk charge of the form $\sin(\pi x/a)/x$ (at 454.8 °C, for the chosen parameters), where $x$ is the radial coordinate.

The application of Arrhenius analysis, summarized in Table I, provides clear signatures for isolated release peaks. Alternatively, the kinetics may be established by fitting (e.g., Levenberg-Marquardt algorithm) a single, full release ramp, with each of the three models, in turn. This approach is mathematically equivalent to the Arrhenius analysis, but it may not converge if started too far (say, $E_a$ off by 5% or more) from the solution (taking the absolute value of $E_a$ leads to convergence in most cases with up to 50% initial deviation). The Arrhenius analysis may prove helpful in this regard, as it provides an unambiguous starting point for the parameter optimization.

Table I.
Effect of a partial linear ramp gas release on the Arrhenius analysis of a second one, for a temperature schedule as in Fig. 1.

| Process | Peak temp. | Arrh. slope | Arrh. intercept |
|---|---|---|---|
| **SD-1** | unchanged | unchanged | unchanged |
| **SD-n, n>1** | higher | unchanged | lower |
| **BLR** | higher | non-linear, no meaningful fit | |

In practice, overlapping or interacting processes are seen, most often. In these cases, two or more ramps can be used to shed light from the peak temperature shifts, as shown in Fig. 3. In particular, a release curve, which does not shift upon partial release ramps, should contain first order desorption components, only, whereas peaks which rapidly approach a limiting temperature should correspond to bulk diffusion limited release.

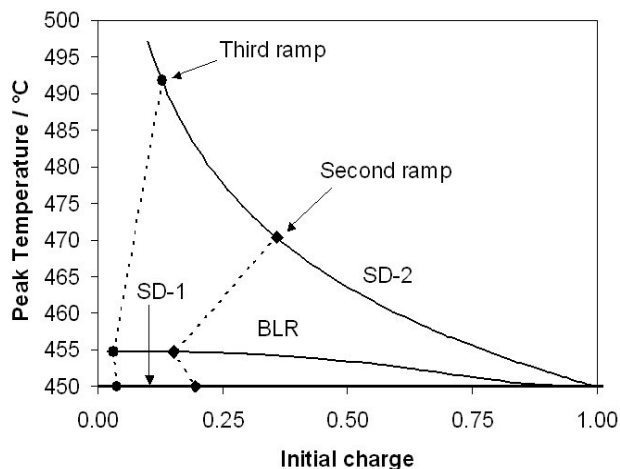

**Figure 3**. Dependence of peak temperature on charge left after a previous release process. First ramp peak set to 450 °C for the three models. Peak temperature of SD-1 process not affected. Dots mark the effect of previous ramps interrupted when the release rate has dropped by 5% from the peak value.

The method can be extended to other sample geometries, as long as an expression can be written for the released gas charge [cf. eq. (7)] as a function of time. These can be obtained by integration of the corresponding solution of the diffusion equation (as can be found[1], for several geometries) over the sample volume.

Porosity, internal to the solid sphere has not been considered. In the case of a porous material, application of these methods would yield effective diffusion parameters which would be relevant to that particular porosity structure.

**Acknowledgment**

This work has been supported by Fondecyt project N°1040213

**References**

[1] Conduction of heat in solids, H. S. Carslaw and J. C. Jaeger, 2nd edition, Oxford University Press, Oxford, 1959.
[2] Principles of adsorption and reactions on solid surfaces, R. I. Masel, J. Wiley & Sons, Inc., New York, 1996
[3] P. Bertone, J. Nucl. Mater. 151, 281 ((1988).
[4] P. A. Redhead, Vacuum 12, 203 (1962).
[5] R. Chen, Surf. Sci. 400, 258 (1998).
[6] A. Halperin, A. A. Braner, A. Ben-Zvi, and N. Kristianpoller, Phys. Rev. 117, 416 (1960).
[7] D. A. King, Surf. Sci. 47, 384 (1975).
[8] J. W. Niemansverdriet, P. Dolle, K. Markert, and K. Wandelt, J. Vac. Sci. Technol. A A5, 875 (1987).
[9] A. L. Cabrera, J. Chem. Phys. 93, 2854 (1990).
[10] R. E. Avila, Jpn. J. Appl. Phys., 43, 7205 (2004).